\documentclass{ws-procs975x65}

\usepackage{amsmath}
\usepackage{amsfonts}
\usepackage{amssymb}
\usepackage{dsfont}
\usepackage{latexsym}

\begin{document}

\title{On the geometrization of the electro-magnetic interaction for a spinning particle}

\author{Francesco Cianfrani$^*$, Irene Milillo$^*$, Giovanni Montani$^{*\dag}$}

\address{$^*$ICRA---International Center for Relativistic Astrophysics\\
Dipartimento di Fisica (G9), Universit\`a  di Roma, ``La Sapienza",\\
Piazzale Aldo Moro 5, 00185 Rome, Italy.}

\address{$^\dag$ENEA-C.R. Frascati (U.T.S. Fusione),\\ via Enrico Fermi 45, 00044 Frascati, Rome, Italy.}

\address{E-mail: francesco.cianfrani@icra.it\\ montani@icra.it}

\begin{abstract}
We outline that, in a Kaluza-Klein framework, not only the electro-magnetic field can be geometrized, but also the dynamics of a charged spinning particle can be inferred from the motion in a 5-dimensional space-time. This result is achieved by the dimensional splitting of Papapetrou equations and by proper identifications of 4-dimensional quantities.\end{abstract}

\keywords{Kaluza-Klein theories.}

\bodymatter

\section*{}
After Einstein recognized the gravitational field as the metric of the space-time manifold, Kaluza and Klein proposed a model in which also the electromagnetic interaction is a geometrical one \cite{K1,K2}. This result has been obtained by adding a spatial closed dimension: the new available five degrees of freedom can be recast as a gauge vector field $A_{\mu}$ and a scalar one $\phi$, under a proper restriction of the general covariance. In particular, the form of the Kaluza-Klein metric tensor is the following one
\begin{equation}
\label{c1}
j_{AB}=\left(\begin{array}{cc}g_{\mu\nu}(x^{\rho})+e^{2}k^{2}\phi^2A_{\mu}(x^{\rho})A_{\nu}(x^{\rho})
& ek\phi^2A_{\mu}(x^{\rho}) \\
ek\phi^2A_{\nu}(x^{\rho})
& \phi^2(x^\rho)\end{array}\right)
\end{equation}  
where Greek letters refer to the standard 4-dimensional space-time coordinates ($\mu=0,\ldots,3$), $e$ is the electron charge and $k$ a constant, while $g_{\mu\nu}$ is the 4-dimensional metric tensor.\\
Hence, by the dimensional reduction of Einstein-Hilbert action, one sees that the Lagrangian for the vector field is the Maxwell one. For what concern the scalar field, it determines the size of the extra-dimension; at the same time, it appears in front of the Maxwell Lagrangian density, so being related to the electro-magnetic coupling constant. Therefore, the stabilization of the additional space corresponds to have a constant electric charge, thus standard electrodynamics has to arise.\\ 
However, it is not enough in view of the geometrization, since also the interaction with matter has to be predicted from the same hypothesis.\\ 
The simplest case is that of a test particle: it follows a geodesics trajectory in the 5-dimensional space-time. One can easily show \cite{cmm1} that the covariant fifth component of the velocity, $u_5$, is a conserved quantity and that, in a 4-dimensional perspective, the motion is that of a test particle endowed with a charge proportional $q$ to the 5-momentum $mu_5$, i.e. $mu_5=q/(2\sqrt{G})$.\\
Moreover, because of the closed nature of the extra-space, the charge is quantized; by imposing its minimum value to be the electron one, an estimate for the length $L$ of the fifth dimension comes out as $L\approx10^{-31}cm$. Being its length just a few order of magnitude greater than Planck's length, we expect to be able to explain the stabilization of the extra-space in a quantum gravity framework.\\
A key-point in the derivation is the link between the fifth- and the fourth-dimensional line elements, i.e.
\begin{eqnarray}
{}^{(5)}\!d s=ds\sqrt{1-u_5^2}
\end{eqnarray}
which implies $|u_{5}|<1\Rightarrow\frac{q}{m}<2\sqrt{G}$, so that the geometrization stands only for macroscopic objects and not for elementary particles.\\
The next step is a rotating body: being $\Sigma^{AB}$ the spin tensor, its dynamics is described by the following system of equations (Papapetrou equations with Pirani condition) \cite{1,3}
\begin{eqnarray}
\left\{\begin{array}{c}\frac{D}{{}^{(5)}\!Ds}{}^{(5)}\!P^{A}=\frac{1}{2}{}^{(5)}\!R_{BCD}^{\phantom1\phantom2\phantom3 A} \Sigma^{BC}{}^{(5)}\!u^{D}\quad\\\\
\frac{D}{{}^{(5)}\!Ds}\Sigma^{AB}={}^{(5)}\!P^{A}{}^{(5)}\!u^{B}-{}^{(5)}\!P^{B}{}^{(5)}\!u^{A}\\\\
{}^{(5)}\!P^{A}={}^{(5)}\!m{}^{(5)}\!u^{A}-\frac{D\Sigma^{AB}}{{}^{(5)}\!Ds}{}^{(5)}\!u_{B}\quad\\\\
\Sigma^{AB}{}^{(5)}\!u_{B}=0\label{pe5}\qquad\qquad\qquad\qquad\end{array}\right..\label{sys}
\end{eqnarray}

While the controvariant 4-dimensional components $\Sigma^{\mu\nu}=S^{\mu\nu}$ can be identified with the 4-spin tensor, the additional components $\Sigma_{5\mu}\equiv S_\mu$ determines a vector, whose physical interpretation is one of the subject of our investigation.\\
By the dimensional reduction of the system (\ref{sys}), we obtain the following one \cite{cmm2}
\begin{eqnarray}
\left\{\begin{array}{c}\frac{D}{Ds}\hat{P}^{\mu}=\frac{1}{2}R_{\alpha\beta\gamma}^{\phantom1\phantom2\phantom3\mu}S^{\alpha\beta}u^{\gamma}+qF^{\mu}_{\phantom1\nu}u^{\nu}+\frac{1}{2}\nabla^{\mu}F^{\nu\rho}M_{\nu\rho}.\\\\
\frac{DS^{\mu\nu}}{Ds}=\hat{P}^{\mu}u^{\nu}-\hat{P}^{\nu}u^{\mu}+F^{\mu}_{\phantom1\rho}M^{\rho\nu}-F^{\nu}_{\phantom1\rho}M^{\rho\mu}
\\\\\frac{D}{Ds}(\alpha^{2}\widetilde{P}_{5}+\frac{1}{4}ekF_{\mu\nu}S^{\mu\nu})=\frac{D}{Ds}q=0
\\\\
\hat{P}^{\mu}=\alpha^{2}P^{\mu}+u_{5}\frac{DS^{\mu}}{Ds}-ekF_{\rho\nu}u^{\rho}S^{\nu\mu}u_{5}+\frac{1}{2}ekF^{\mu}_{\phantom1\rho}S^{\rho}\\\\
S^{\nu\mu}u_{\nu}+S^{\mu}u_{5}=0\label{pe5s}\end{array}\right.,\label{sys2}
\end{eqnarray}
where for the quantity $M^{\mu\nu}$ we have
\begin{eqnarray}
M^{\mu\nu}=\frac{1}{2}ek(S^{\mu\nu}u_{5}+u^{\mu}S^{\nu}-u^{\nu}S^{\mu})\label{emmom}.
\end{eqnarray} 
Once we think at the quantity $q$ as the electric charge of the system (a strong indication for this comes from the third equation of the system (\ref{sys2}), which tells us it is conserved during the motion) we find that the first two equations coincide with those describing the dynamics of a rotating body with an electromagnetic moment $M^{\mu\nu}$ (Dixon-Souriau equations \cite{5,4}) . Furthermore, it is clear that the additional components of the spin tensor describe a non-vanishing electric moment, since the vector $S^\mu$ enters into the electro-magnetic moment with a term proportional to the velocity. In fact, in a co-moving frame the spatial part of the electro-magnetic moment, i.e. $M^{ij}$, receive no contribution from terms with $S^{\mu}$.\\ 
Therefore, a rotating body in a Kaluza-Klein background behaves as a charge rotating particle in a 4-dimensional point of view. A proper feature of such models is an electric dipole moments, associated with additional components of the spin tensor.\\
It arises the question of the possibility to implement this scheme in a quantum framework, because an electric moment for elementary particles implies the violation of the parity and of the time-reversal invariance.\\ 
We want to emphasize that, while Kaluza-Klein theories preserve both P and T, definitions of parity and time-reversal on 5-dimensional spinors differ from those on 4-dimensional ones, so violations of the latter do not imply violations of the former.\\
For example, since in five dimensions the $\gamma_5$ matrix is one of Dirac matrices, an explicitly 4-parity-violating term appears in the Dirac action, while the representation of the 5-parity is given by $i\gamma^0\gamma_5$ and it is conserved.\\ 
For what concern the time-reversal, the question is more subtle, however its violation is not surprising, since an electric dipole moment term arises directly from spinor connections in five dimensions \cite{edm}.\\

\end{document}